# Anti-Black racism workshop during the Vera C. Rubin Observatory virtual 2021 Project and Community Workshop

**Andrés A. Plazas Malagón***
Kavli Institute for Particle Astrophysics and Cosmology, P.O. Box 20450, MS29, Stanford, CA 94309, USA
SLAC National Accelerator Laboratory, 2575 Sand Hill Road, MS29, Menlo Park, CA  94025, USA
Department of Astrophysical Sciences, Princeton University, Peyton Hall, Princeton, NJ 08544, USA

**Federica Bianco**
University of Delaware Department of Physics and Astronomy 217 Sharp Lab Newark, DE 19716 USA
University of Delaware Joseph R. Biden, Jr. School of Public Policy and Administration, 184 Academy St, Newark, DE 19716 USA
University of Delaware Data Science Institute
Center for Urban Science and Progress, New York University, 370 Jay St, Brooklyn, NY 11201, USA

**Ranpal Gill**
Rubin Observatory
NSF's NOIRLab, 950 N Cherry Ave, Tucson, AZ 85719, USA

**Robert D. Blum**
Rubin Observatory
NSF's NOIRLab, 950 N Cherry Ave, Tucson, AZ 85719, USA

**Rosaria (Sara) Bonito**
INAF - Osservatorio Astronomico di Palermo, Piazza del Parlamento 1, 90134 Palermo, Italy

**Wil O'Mullane**
Rubin Observatory
NSF's NOIRLab, 950 N Cherry Ave, Tucson, AZ 85719, USA

**Alsyha Shugart**
Rubin Observatory
NSF's NOIRLab, 950 N Cherry Ave, Tucson, AZ 85719, USA

**Rachel Street**
Las Cumbres Observatory, 6740 Cortona Drive, Suite 102, Goleta, CA 93117, USA

**Aprajita Verma**
Sub-department of Astrophysics, University of Oxford, Denys Wilkinson Building, Keble Road, Oxford OX1 3RH, UK

**\*Corresponding author:** plazas@slac.stanford.edu

**Abstract**

Systemic racism is a ubiquitous theme in societies worldwide and plays a central role in shaping our economic, social, and academic institutions. The Vera C. Rubin Observatory is a major US ground-based facility based in Chile with international participation. The Observatory is an example of excellence and will deliver the largest survey of the sky ever attempted. Rubin's full scientific and social potential can not be attained without addressing systemic racism and associated barriers to equity, diversity, and inclusion (EDI). During Rubin's 2021 virtual Project and Community Workshop (PCW), the annual Rubin community based meeting, an anti-Black racism workshop took place, facilitated by "The BIPOC Project" organization. About 60 members from different parts of the Rubin ecosystem participated. We describe the motivation, organization, challenges, outcomes, and near- and long-term goals of this workshop.

**Introduction**

The Vera C. Rubin Observatory (Rubin) was ranked as the highest priority ground-based astronomical facility in the 2010 National Academies of Sciences, Engineering, and Medicine Decadal Survey of Astronomy and Astrophysics (Astro2010)[1]. The same report recognized the lack of progress in increasing the number of minorities in astronomy and the work that is still needed to eliminate gender gaps in our field. A decade later, the next iteration of this decadal report[2] placed a more explicit emphasis in how Equity, Diversity, and Inclusion (EDI) are fundamental for the foundations of the profession. Understanding racism as structural and systemic, rather than being an individual and intentional occurrence, can prompt organizations to confront the root causes of racial disparities and lack of progress with EDI efforts. EDI is important for the Rubin Observatory not just on an ethical level, but also for the potential to maximize the science return of the project by (among other things) broadening the talent pool.[3] During Rubin's 2021 virtual Project and Community Workshop (PCW), the annual Rubin community based meeting, we proposed the realization of an anti-Black racism workshop facilitated by an external organization (*"The BIPOC Project"*)[4]. The initial plan proposed having a workshop on systemic racism and its relation to EDI, however, triggered by the events that took place in 2020 in the USA and at an international level in the aftermath of the murder of George

---

[1] [New Worlds, New Horizons in Astronomy and Astrophysics](#)
[2] [Pathways to Discovery in Astronomy and Astrophysics for the 2020s](#)
[3] See, for example, Page, S. E. "The Difference: How the Power of Diversity Creates Better Groups, Firms, Schools, and Societies." 2008, and Nature editorial "Science benefits from diversity", 2018.
[4] [https://www.thebipocproject.org/](https://www.thebipocproject.org/)

Floyd, the Black Lives Matter movement and the Strike for Black Lives on 19 June 2020[5], the final workshop was more specifically focused to address anti-Black racism. Within what we refer to as the "Rubin Ecosystem"[6], it is clear that there is work to do to improve EDI and to address and dismantle systemic racism, in particular, anti-Black racism. The number of Black and African American people in the scientific and technical Rubin community is unlikely to surpass 1% if attention is not paid to this concern Currently ,there are no official metrics or statistics to track such Rubin workforce demographics with any accuracy; this is itself a problem. To create a community that is naturally inclusive and equitable, sustainable systems to achieve that goal need to be implemented to ensure that members of groups traditionally marginalized from STEM (Science, Technology, Engineering, and Math) and astronomy feel welcome and supported, thus increasing their retention in astronomy. The goal of the workshop was to educate a representative group of Rubin members, including senior leadership and people in decision-making positions, across the Rubin organization so that inroads could be made to put in place systems and processes that lead to positive change. After the virtual workshop participants were expected to "return" to their institution/collaboration and find ways to make actionable change for incremental progress that would eventually address the underrepresentation of Black people in organizations associated with Rubin Observatory. Given the predominant role of Rubin as one of the most important and visible astronomical projects of the decade, we hope that this also sets an example in the astronomical community in general.

**The Rubin Ecosystem and structure of the workshop**

The Rubin Observatory is a large organization. The Rubin Observatory includes several hundred employees that work on various aspects of Rubin Construction Project and Rubin Operations. These are separate, although overlapping, organizations with teams that focus on different areas of the project such as Data Management, Commissioning, etc. In total we identified 8 subsystems within the Rubin Ecosystem. This staff includes Rubin's Chilean employees working in both the US and Chile.

There is also the community that is interested in generating science from the upcoming data from Rubin's Legacy Survey of Space and Time (LSST). This includes thousands of people and many of these researchers are organized into the eight LSST Science Collaborations, which are supported by the LSST Corporation , a private non-profit engaging in fund-raising. Both these organizations are external to Rubin, but they are key elements of the Rubin community and ecosystem.

A typical annual meeting, called the Project Community workshop (PCW), includes more than 500 participants from all parts of the Rubin EcoSystem.

---

[5] https://m4bl.org/events/strike-for-black-lives/ , https://www.particlesforjustice.org/, https://www.shutdownstem.com/

[6] The Rubin ecosystem is formed by funding agencies and institutions, the Rubin Construction Project (formerly known as LSST Construction project) which includes engineers, scientists who will operate the telescopes, international collaborations of scientists, etc. A detailed structural map can be found here. https://noirlab.edu/public/images/rubin-rubin-ecosystem/

Workshop Planning

Initiatives aimed at improving climate and equity from the inside, such as committees and discussion groups, are common both in academic STEM fields and in Rubin[7]. However, these initiatives, as well- intentioned as they are, often lack strategic goals and risk being slow or even completely ineffective at enacting change. With many equity issues to address in academic STEM, in most cases, these groups have little time and, for the most part, no resources for (self)-education and training. We decided that education was key to enact change and planned to recruit a professional team to facilitate training. The implementation of this required securing funds: the cost of such training is not trivial, typically running in the several thousands of dollars. We took advantage of a regular funding call for proposals issued by the LSST Corporation. We submitted a proposal to the LSSTC 2020 Enabling Science call, expecting to run in-person anti-racism training sessions at the 2021 Project Community Workshop, but when the PCW 2020 was moved to virtual format, the final decision was made to conduct the workshop remotely.

To plan the proposal we had contacted an organization ("*Crossroads*"[8]) that at the time organized anti-racism training for academic communities and based the budget on their quote. The timing was peculiar: our proposal was submitted before the start of events that catalyzed anti-racism and anti-Black racism, especially in the US. During the US "racial reckoning" triggered by the COVID-19 pandemic and the murder of George Floyd, organizations such as *Crossroads* –the organization we had initially contacted– and *The BIPOC Project*[9] –the organization we finally selected– saw a spike in requests. This led to an increase in the cost of this facilitated workshop, and more competition to recruit a facilitators team for the dates of the PCW, which were fixed. Several organizations, including *Crossroads,* had by 2021 reorganized their training offerings and were no longer providing small workshops such as the one we had planned. The cost also increased but the Rubin Observatory Leadership agreed to contribute to the expenses. However, the changing political landscape of 2020 meant that we had to be aware of the ability to use federal funding for the activity.[10] Ultimately, with a change of federal administration, we were able to use some federal funding.[11]

Selection of Facilitating organization

In order to identify an organization to facilitate this workshop, we initially consulted with colleagues in different institutions and performed our own research in the literature and the

---

[7] Thomas+2022: Creating an inclusive and diverse environment at Vera C. Rubin Observatory
[8] https://crossroadsantiracism.org/
[9] https://www.thebipocproject.org/ Other organizations contacted included "Overcoming Racism" (https://www.overcomeracism.com/) and "Diversity Talks" (https://www.diversitytalkspd.com/)
[10] https://www.npr.org/2020/09/05/910053496/trump-tells-agencies-to-end-trainings-on-white-privilege-and-critical-race-theor
[11] https://law.ucla.edu/news/biden-reverses-trump-executive-order-banning-diversity-training

internet in general. In the end, when selecting the facilitating organization we considered the following:

- **Cost**: we had a hard limit to the budget. We made that clear to the organizations we contacted from the start.
- **Timing**: our sessions were tied to the dates of the Rubin PCW with no flexibility beyond the PCW week. Furthermore, the meeting was offered online to an international community with attendees from as far west as Hawai'i and as far east as East Asia and Oceania the live time was restricted to 10AM-4PM Eastern (USA) to limit the strain on participants from non-US time zones.
- **Focus on anti-racism over broader equity and EDI topics** - to ensure effectiveness we wanted to ensure that this workshop was focused and did not generalize EDI issues. We had already chosen the focus to be anti-racism and further narrowed the scope to be anti-Black racism.
- Given the scope, we decided it was critical that the **leadership of the facilitating organization included Black, Indigenous, People of Color (BIPOC) and that BIPOC facilitators would be visible to the participants**. This helped us to ensure that the recommendations we would be receiving would come from personal experience and knowledge, not just from academic research. The choice also allowed us to support a BIPOC owned and run business.
- We ensured that the organizations had **experience with academic environments, ideally STEM**. Given the cost, these facilitated workshops are more commonly held in the corporate world than in academia. We believe that those in academia are, in general, often exposed to equity concerns and issues, yet inequities, biases, and discrimination are still pervasive, and even more so in STEM. In addition, we acknowledge that STEM training and the traditional STEM environments tend to be competitive and even aggressive; while EDI issues are more significant in STEM than in other Academic environments, the personalities in the STEM networks tend to be analytical and skeptical of outsiders' approaches.
- Rubin is an international community and this exposes a significant issue in EDI work: equity, inclusivity, and racism issues may look different in different cultural contexts.[12] **We asked the organizations to describe how they would approach training with an international audience**. Ideally we wanted to ensure that the organization had prior international experience, but we found that very few such organizations work across international boundaries. We relied on a commitment of the organizers to prepare material that explicitly supported a multi-cultural, international audience.

During the selection process, we had several phone and video calls with representatives of the following candidate organizations: *Crossroads, Overcoming Racism, Diversity Talks, and The BIPOC Project.* Based on the criteria and constraints described above, in the end we selected *"The BIPOC Project"* to facilitate the following workshop at Rubin's 2021 virtual PCW[13]:

---

[12] In many cultures and countries, darker skin is also a basis for discrimination.
[13] https://www.thebipocproject.org/what-we-do

*"Building Black Power: Dismantling Anti-Blackness in Our Institutions and Movements –
This experiential gathering tackles the root causes of anti-Black racism and its cultural influences within our institutions, movements and communities, and offers a framework for unlearning anti-Blackness and developing a pro-Black stance within anti-racist practice among BIPOC folks. "*

Selection of Participants and Leadership Involvement

The original proposal had envisioned an in-person training opportunity with the expectation to include between 100-150 participants. This number was based on the participation in structured, but not facilitated, EDI sessions held in previous years and on the expected number of participants. We did not initially anticipate having to select participants. However, the modification of the original plan described above and the maximum number of participants set by *The BIPOC Project* to maximize the effectiveness of training (based on their experience with prior events[14]), we were faced with the possibility of having to select participants among applicants. We created a sign-up sheet for the workshop within the PCW registration to assess potential interest.  We indicated in the initial sign up and iterated for clarity by email that we expected full participation to the entire workshop, which was held over two days in parallel with scientific sessions at the PCW. Further, we explicitly recruited representatives of the Rubin communities to ensure the entire Rubin ecosystem would be exposed to this training.  In addition, we recruited members of the Rubin Science Advisory Committee, the official body advising Rubin Observatory on scientific decisions; while the scope of their recommendations are primarily scientific, or science-motivated, we recognize that bias and racism are inextricably linked to our way of thinking and reflecting on the world, including our scientific thinking and science-driven prioritization.

Finally, we wanted to ensure participation by the organizations that manage and govern Rubin. Change can come from a grassroots level as has mostly been the case with the Rubin Observatory EDI initiatives. It is recognized that commitment from leadership sends a strong message to the rest of the organization. While the workshop was oversubscribed according to the interest expressed during registration to the PCW, a number of Rubin Leaders, both current and future, were explicitly invited to participate and approximately 80% attended. It was invaluable to have leadership involved because it allowed subsequent conversations to take place with those individuals using language, terms, and ideas that were learned in the workshop. We invited representatives from Rubin's main funding agencies in the USA, the National Science Foundation (NSF), and the US Department of Energy (DOE),  as well as the leadership of the organizations that manage Rubin: the Association of Universities for Research in Astronomy (AURA), the NSF's National Optical-Infrared Astronomy Research Laboratory (NOIRLab) and DOE's SLAC National Accelerator Laboratory (SLAC).

In total we received 74 applications and obtained permission from *The BIPOC Project* to extend the participants' limit to this number so that we did not have to reject any applicants.

---

[14] *The BIPOC Project* organizes workshops for up to 60 Participants with two two-hour sessions over consecutive days.

We monitored the demographic profile of our participants by asking questions in the application form including nationality and their seniority level, defined in three tiers: junior (students and postdoctoral scholars), mid-career (senior postdocs, scientists, junior faculty), and senior (faculty starting at associate level, senior scientists). We did not ask participants to identify their gender or ethnicity/race. In retrospect we realized monitoring the fraction of People of Color (POC) signed up for the workshop was particularly important. The organizers browsed the list of participants to provide additional race or ethnicity information for those whom they had prior knowledge and assigned them to two groups: POC[15] and non-POC in order to facilitate broad discussion in small groups.

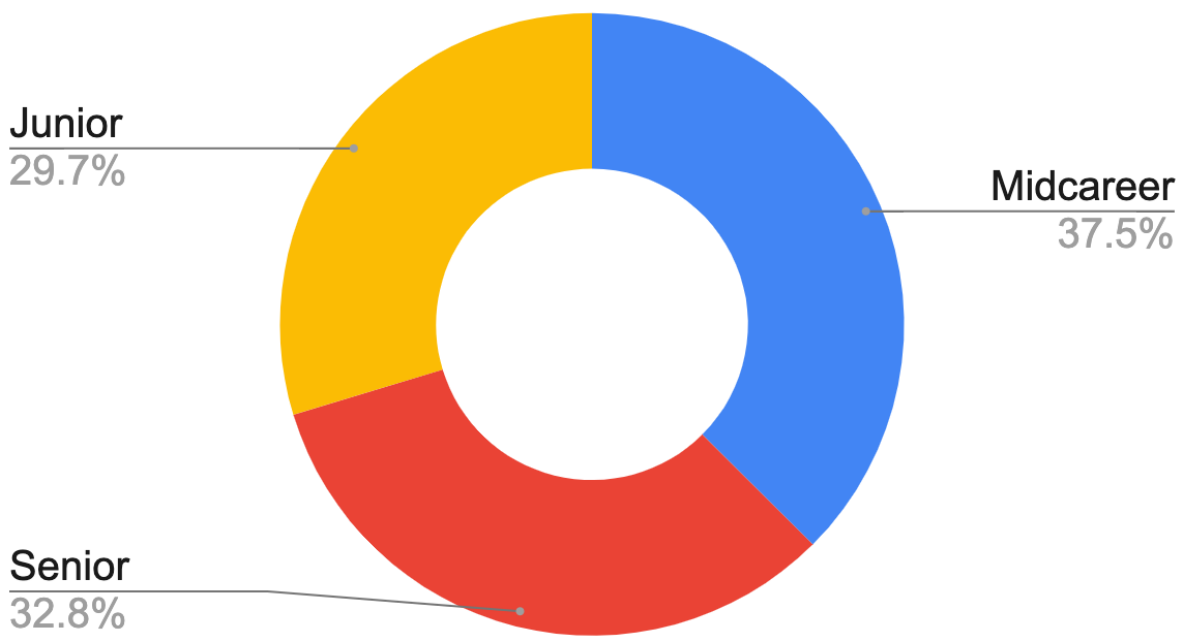

---

[15] We defined Person of Color (POC) as anyone who is not white.

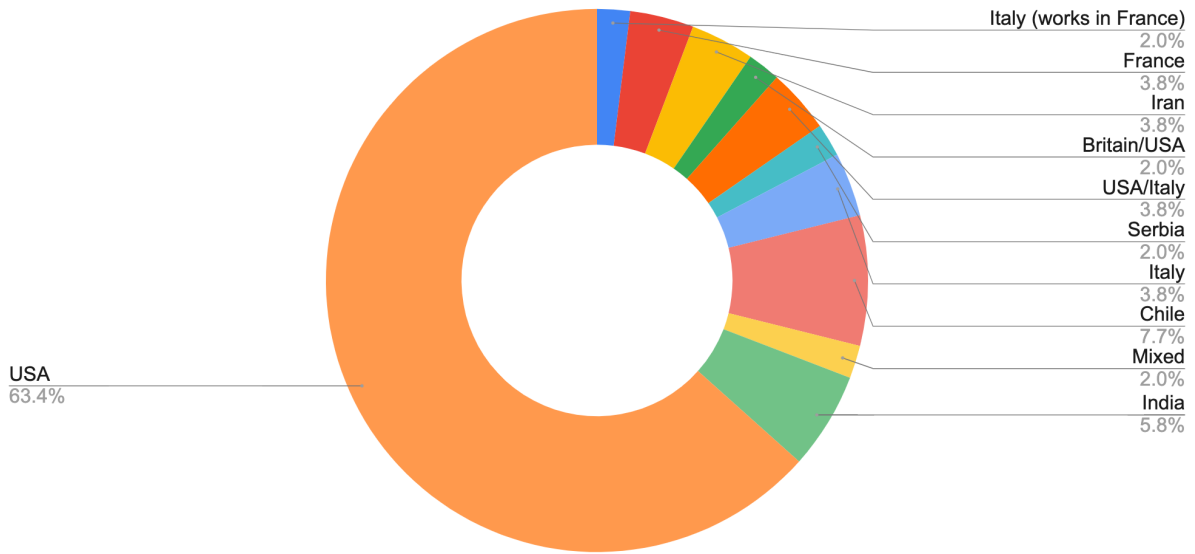

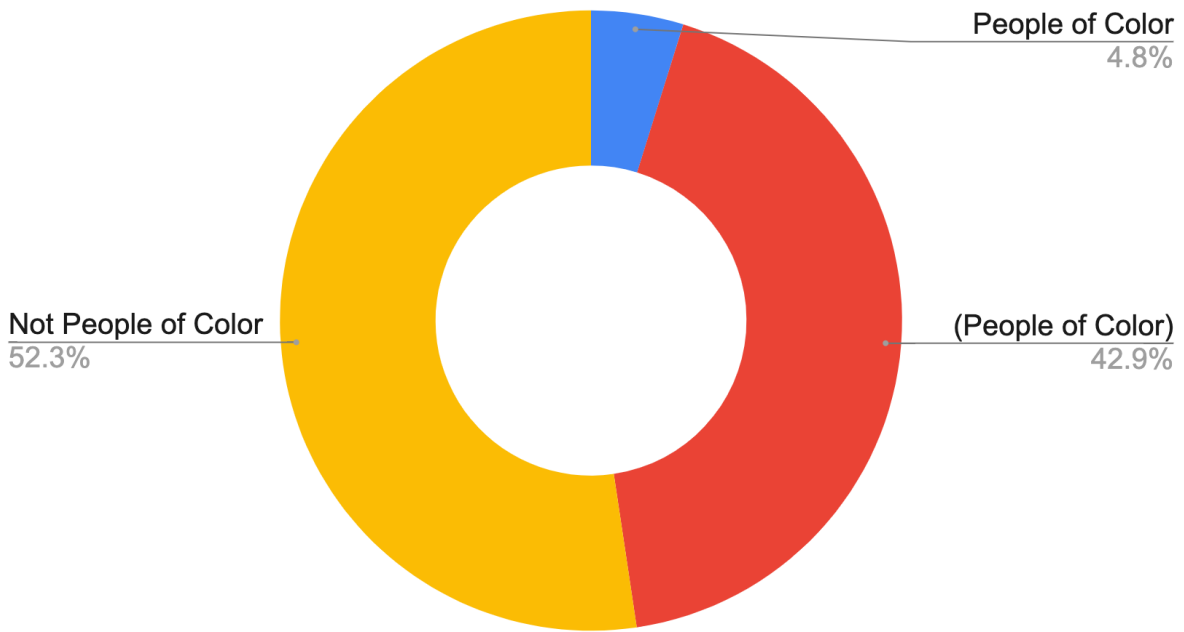

Figure Captions
*Top:* Participants seniority (%): junior (red), mid-career (yellow), senior (blue)
*Middle:* Participants nationality(%)

*Bottom:* Fraction of participants who were People of Color: People of Color (blue and red), Not People of Color (red). When in parenthesis, the fraction represents an assumed identity by the organizers based on personal knowledge of the participants

**Workshop sessions**

Both sessions of the two day workshop were facilitated by *The BIPOC Project*. The first session consisted of an introduction to the workshop and to fundamental concepts, definitions, and shared vocabulary. The desired outcomes from the workshop, were as follows:

- *"A shared language and analysis for understanding and addressing white supremacy and anti-Black racism as foundational to advancing diversity, equity, and inclusion within your community"*
- *"Deeper awareness of the ways in which anti-Black racism shows up in ourselves, in society, and in our organizations"*
- *"Initial steps to undo anti-Black racism, advance racial equity, and build a pro-Black scientific future"*

During this session, participants were encouraged to reflect on their personal reasons to engage in anti-racism work within the Rubin Ecosystem, and discuss the definition of racism as a system of oppression and advantage based on race. Emphasis was placed on the distinction between diversity, inclusion, and equity and equality (and their relation to justice), as well as on the importance of performing a race-explicit analysis that names White Supremacy and anti-Black Racism. The second session was more interactive in nature, relying heavily on Zoom breakout rooms in order to overcome the limitations from the virtual format of the workshop. Participants were prompted by the facilitators to identify organizational and individual barriers to building racially equitable and pro-Black futures in the Rubin community, as well as promising practices to address racial inequities and advance opportunities for Black scientists in the Rubin Ecosystem. At the end, participants produced a document with concrete ideas to address anti-Black racism in their institutions and spheres of influence.[16]

**Early outcomes**

Increased awareness of EDI and particularly awareness of systemic disadvantages faced by Black people in the USA and further afield has brought about a movement within Rubin Observatory Ecosystem to take immediate action. The workshop and the realization of the issues that it brought forth helped put a spotlight on systemic racism, anti-Black racism and their relation to EDI. This has resulted in most hiring managers requiring a statement from job applicants about their personal EDI efforts. These are a part of the evaluation process and are considered as important as technical/other skills. This is reinforcing the message that the value and importance of diverse teams is understood and internalized.

---

[16] Document: **https://padlet.com/ELteam/vibxvs1uff440un2**

In addition, several months after the workshop we organized a follow-up session, via Zoom, in which we invited the participants of the workshop to 1) share how they have used the tools from the workshop, 2) share what actions they have taken inspired by the workshop, 3) collect feedback from the workshop, and 4) propose new ideas for next actions and future workshops. 17 of the about 70 workshop participants attended this follow-up session. In what follows, we reproduce particular examples shared by the participants of this follow-up session:

- Construction Project Staff: I found what grabbed my attention from the workshop is the message that if you are not actively trying to seek and dismantle barriers, you are passively upholding them - I try to work within that framework in NOIRLab advocacy - bringing into the light the shared responsibility for increased representation. For example, when the organization published communications about Black History Month some of us found that a lot of the responsibility was put on people of color in our organization to produce these messages. This should not be the case. In addition to communications, we need to push the shared responsibility of addressing barriers in recruitment when serving on hiring committees.

- University Faculty and Science Collaboration Lead: Faculty recruitment took on much of the advice that were raised in the last workshop, including wording and advertising to other audiences. We encountered positive reactions from the University but without directable action to enable them, so we came up with our own strategy to ensure diverse and unbiased recruitment for these positions. There were also difficulties establishing positions that are explicitly designated for individuals holding specific identities due to legal issues. For the faculty search, we had an explicit item in rubric for contribution to EDI. The identity of the candidate can be a direct contribution to the diversity, equity, and inclusion of the department.

- University Faculty and Science Collaboration Member: USCS insists on protocols and the academic personnel office runs training relating to how to review the EDI statements. Yet at UCSC we do not have a Black population at all. We have added a focus group of our diversity and climate committee on Black representation cause based on our population breakdown we just often miss a Black representation .Over the past year we created a safe space in our division specifically aimed at making space for Black students inspired by the American Institute of Physics National Task Force to Elevate African American Representation in Undergraduate Physics & Astronomy (TEAM-UP) report[17], we are taking each of the pieces recommended by the TEAM-UP report and also representing Rubin at National Society of Black Physicists (NSBP) meetings.[18]

- LSSTC Science Staff: The LSST Corporation has run a selection process for postdoc fellows (Catalyst Fellowship), and I leveraged what I learned in the workshop as I was overseeing the selection process. Some info from the workshop helped us pass some of the new practices that we were concerned might get pushed back. The reviewers actually indicated that the anonymization of the process triggered a different kind of perspective. In summary what I learned was really to include EDI considerations in every step of the process, not as an add-on. In the LSSTC fellowship process, it was possible to craft the fellowship in a way that would be more attractive to Black candidates. [We] had the flexibility to create a program and emphasize career support, mentorship provided, and leadership training that would be provided, recognizing multiple ways of benefiting the community. We were able to create a position that would be more distinctive. We also had an application in two parts. Applicants write both a community impact statement (could be personal experience) and a research statement. These are weighted equally. Did not need to explicitly include diversity criteria to achieve a diverse pool of finalists because community impact was built into the process.

---

[17] https://www.aip.org/diversity-initiatives/team-up-task-force
[18] https://nsbp.org/page/2022NSBPConferenceSummary

- University Faculty, Rubin Operations and Science Collaboration Member: After the workshop I took materials from the workshop and dedicated one day in the class in the semester to go over EDI topics. The feedback I got was that they [the students] never have physics faculty members talk about these things or have anyone tell them that being cognisant of these issues is an important part of their professional development and I was struck by how undergrads are not exposed to these topics at all. In the Physics department, I was able to make funds available for an undergraduate physics fellowship for students from underrepresented groups (could not be done for Black students because of legal issues). As I was researching how to set this up, I found out that Hubert Mack Thaxton who I consider to be a significant historical figure did his PhD thesis at our University on proton-proton scattering and we named the fellowship after him. I engaged into researching him to include information about him in the fellowship and his family reached out to me to thank us for this.

- Rubin Leadership Team: Rubin is committed and we have been working to implement some of the things we learned from this workshop, things similar to what our colleagues from the LSSTC referred to: trying to be aware of biases and take concrete steps in recruitment to avoid them. There are times, however, when you, if you are **not** biased, you are not making a difference. You need to be deliberate to include for example Black scientists in job searches or even think of targeted hires as many organizations do for many reasons. We have not been successful at hiring a Black scientist. In the current searches, we have no Black scientists in the applicant pool.

- Rubin Construction/Operations Staff: Another thing we tried is Land Acknowledgement. We tried to include it for a few presentations. It was considered not completely appropriate with its wording. Continuing to work on this. Indigenous members of NOIRLab community are helping with the message.

The participants' experiences can be grouped into several themes. One of these themes was the importance of actively seeking and breaking down barriers, and that the responsibility for increasing representation should not rest solely on individuals from underrepresented groups. Another theme was the need to include EDI considerations in every step of the process, rather than as an add-on. Participants reported success in creating fellowships and positions that were more attractive to Black candidates by emphasizing career support, mentorship, and leadership training, and by creating an application process that weighted community impact equally with research statements. They also described creating safe spaces for Black students, and developing class materials to educate students about EDI topics. Legal issues were noted as barriers to some initiatives, such as creating positions explicitly for individuals with specific identities or funding fellowships for Black students. Some participants reported that despite their efforts, they had not been successful in hiring Black scientists, and suggested that targeted hires might be necessary. Overall, the participants found the workshop to be a valuable tool for advancing EDI efforts in their respective fields.

**Lessons learned and challenges**

- *Be clear about the topic:* Rubin is an international collaboration and equity does look different in different local contexts. While anti-Black racism is demonstrably a global issue, different cultures experience the specificity and criticality of anti-Black racism differently. This resulted in a range of reactions that included participants asking why we were being so specific addressing only anti-Black issues and participants feeling like their specific experiences of marginalization had been overlooked. Nonetheless, the organizers stand by the choice of focusing on a narrow and timely issue which allowed the conversation to be specific and detailed. The organizers' own experience with more general "anti-discrimination" workshops, or even generalized anti-racism work is that broad topics end up in forcing generalization that do not allow for specific planning on effective action. After the first session, one of the participants expressed their frustration about how certain participants kept centering themselves instead of Black people and scientists in the conversation. Ultimately, whatever topic is selected by organizers, we suggest specificity.

- *Ensure commitment to follow-up after the workshop and report back*: It's normal that momentum is lost almost immediately after the workshop is over, so before the first workshop, it is useful to set a check-in schedule, and ask people to commit to sharing their EDI work and perceived outcomes (negative and positive) so that everyone can learn from them. In our case, we initially considered gathering feedback from the workshop participants via a Google form after the workshop. In the end, as indicated above, we organized a follow-up session, via Zoom, in which we invited the participants of the workshop to 1) share how they have used the tools from the workshop, 2) share what actions they have taken inspired by the workshop, 3) collect feedback from the workshop, and 4) propose new ideas for next actions and future workshops.

- *Collect data to measure impact*: It can be difficult to measure success. Whether someone feels a sense of community or belonging are quite intangible. Other metrics can be quite concrete such as measuring over time if the number of Black job applicants has changed, keeping track of the number of subgroups or ecosystem members involved in proactive EDI activities, and recording the number and kinds of EDI-related activities that result after the event to compare with those before. However, attributing such a change back to actions taken by the organization or its individuals is challenging. In addition, it is also challenging that some organizations do not allow such data to be collected and even if it is, Human Resources departments are reluctant to share it. Insight can also be gained by hiring social scientists or other professionals who can inform on how to gather these metrics properly, perhaps via hiring organizations to conduct climate surveys or other mechanisms, and who can provide suggestions on how to act based on the findings.

- It was not clear if involving Black and African American Rubin members in the workshop resulted in any benefit. Some of the Black and African American participants reported feeling that the workshop was not effective for them and instead it put them in the position of having to hold participants' hands through realization of a reality that is all too obvious to them. Conversely, we wonder if a workshop on anti-Black racism might have felt more abstract and academic if the participants did not have the Black members of this community under their eyes at the time of the workshop. If minority colleagues are included or asked to be included, monetary compensation or compensation in the form of enhanced job-security should be offered.[19] In addition, analogously to "affinity spaces", "solidarity spaces" could be created for white people only to educate themselves and ask learning questions without burdening their minority colleagues.[20]

**Acknowledgements**

We thank Dara Norman, Dorian Russell, and Tim Sacco for the invitation to write this contribution and for editorial and content suggestions that improved the manuscript, Brian Nord for important feedback and discussions during the writing of the LSSTC Enabling Science 2020 proposal that partially funded the workshop, the LSST Corporation and the leadership of the Rubin Observatory for funding the workshop, Las Cumbres Observatory for administration of the LSSTC grant, the workshop and follow-up session participants, *The BIPOC Project*, Fiona Kanagasingam and Merle Mcgee for facilitating the workshop, the organizers of the virtual Rubin Project and Community Workshop 2021, Anissa Tanweer for feedback on the manuscript, and the additional co-PIs of the LSSTC proposal for their support (Amanda Bauer, Jeffrey D. Barr, William Brandt, Patricia Burchat, Rachel Mandelbaum, Brian Nord, Chad Schafer, Sandrine Thomas). The work of AAPM was supported by the U.S. Department of Energy under contract number DE-AC02-76SF00515.

---

[19] Suggestion by B. Nord (Fermilab), in the context of the Dark Energy Survey project.
[20] Suggestion by T. Jackson during their plenary talk in the virtual AIP TEAM-UP Implementation Workshop of January 2021: https://www.aip.org/diversity-initiatives/teamup-implementation-workshop-1